\documentclass[prl,twocolumn,superscriptaddress]{revtex4-1}

\usepackage{amsmath,bm,amsfonts,graphicx}

\newcommand{\ket}[1]{\vert#1\rangle}
\newcommand{\bra}[1]{\langle#1\vert}

\begin{document}

\title{Edge theories in Projected Entangled Pair State models}

\newcommand{\MPQ}{\address{Max-Planck-Institut f\"ur Quantenoptik, Hans-Kopfermann-Str.\ 1, D-85748 Garching, Germany}}
\newcommand{\RWTH}{\address{Institut f\"ur Quanteninformation, RWTH Aachen University, D-52056 Aachen, Germany}}
\newcommand{\Gent}{\address{Department of Physics and Astronomy,  Ghent University, Ghent, Belgium}}
\newcommand{\Vienna}{\address{Vienna Center for Quantum Science and Technology, Universit\"at Wien, Boltzmanngasse 5, A-1090 Wien, Austria}}
\newcommand{\Toulouse}{\address{Laboratoire de Physique Th\'eorique, C.N.R.S.\ and Universit\'e de Toulouse, F-31062 Toulouse, France}}

\author{S.~Yang}\MPQ
\author{L.~Lehman}\RWTH
\author{D.~Poilblanc}\Toulouse
\author{K.~Van~Acoleyen}\Gent
\author{F.~Verstraete}\Gent\Vienna
\author{J.I.~Cirac}\MPQ
\author{N.~Schuch}\RWTH

\begin{abstract}
We study the edge physics of gapped quantum systems in the framework of
Projected Entangled Pair State (PEPS) models.  We show that the effective
low-energy model for any region acts on the entanglement degrees of freedom at
the boundary, corresponding to physical excitations located at the edge.
This allows us to determine the edge Hamiltonian in the vicinity of PEPS
models, and we demonstrate that by choosing the appropriate bulk
perturbation, the edge Hamiltonian can exhibit a rich phase diagram and
phase transitions.  While for models in the trivial phase any Hamiltonian
can be realized at the edge, we show that for topological models, the edge
Hamiltonian is constrained by the topological order in the bulk which can
e.g.\ protect a ferromagnetic Ising chain at the edge against spontaneous
symmetry breaking.
\end{abstract}

\maketitle

The edge of strongly correlated quantum systems can display very
intriguing phenomena. For instance, in two-dimensional (2D) quantum Hall
systems the low energy behavior can be described in terms of chiral modes
which live at the edge of the
material~\cite{wen:fqhe-edge,wen:fqhe-edge-review}.  Interestingly, these edge
modes cannot be described by a conventional one-dimensional (1D) theory,
and their properties are dictated by the presence of a topologically
ordered bulk.

In this Letter, we study the low-energy physics for a class of spin
systems on 2D lattices.  We show that the Hilbert space of the effective
low-energy theory can be identified with the entanglement degrees of
freedom which live at the edge of the system.  This allows us to construct
1D edge Hamiltonians which describe the low-energy physics of the system,
and investigate how they change under perturbations in the bulk.  We find
that bulk perturbations can induce phase transitions at the boundary, and
explicitly investigate one particular example where we find a rich phase
diagram with gapped, gapless, and symmetry-broken phases at the boundary.
We also study the effect of topological order in the bulk and find that it
induces constraints on the edge Hamiltonian which cannot occur in
conventional 1D spin systems, a direct consequence of the topological
protection~\cite{gu:spt-fermions,chen:spt-bosons};  for instance, we give
a model based on the Toric Code~\cite{kitaev:toriccode} whose edge
Hamiltonian is an Ising chain, but which is protected against spontaneous
symmetry breaking by the topological properties of the bulk.

We restrict our attention to Projected Entangled Pair State (PEPS) models,
and perturbations thereof.  PEPS models consist of a Hamiltonian
$H$ together with its ground space which are both derived from a single
tensor which describes the entanglement structure of the system
locally~\cite{aklt,verstraete:mbc-peps,perez-garcia:parent-ham-2d,schuch:peps-sym}.
We focus on models where $H=\sum h$ is translational invariant,
i.e., a sum of identical local terms, and gapped for periodic
boundaries (PBC).  Many paradigmatic models such as the AKLT
model~\cite{aklt}, topologically ordered
systems~\cite{verstraete:comp-power-of-peps,buerschaper:stringnet-peps,gu:stringnet-peps},
or Resonating Valence Bond (RVB)
states~\cite{verstraete:comp-power-of-peps,schuch:rvb-kagome} are PEPS
models; and we will illustrate our results with particular perturbations
of these models.

\begin{figure}[b]
\includegraphics[width=\columnwidth]{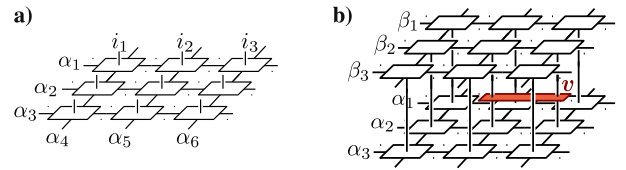}
\caption{\label{fig:peps}
\textbf{a)} Construction of a PEPS by contracting local tensors.   PEPS
give a map from the boundary indices
$(\alpha_1,\alpha_2,\dots)$ to the bulk indices $(i_1,i_2,\dots)$.
\textbf{b)} Using this map, any bulk Hamiltonian $v$ naturally induces a
Hamiltonian on the boundary by sandwiching $v$ in between the
PEPS. Note that the boundary degrees of freedom still need to be
orthogonalized.}
\end{figure}

We start by introducing PEPS models.  For simplicity, we restrict to
square lattices and translationally invariant systems. The central object
is a five-index tensor $A^i_{\mu_1,\mu_2,\mu_3,\mu_4}$, with
\emph{physical index} $i=1,\dots,d$ and \emph{virtual indices}
$\mu_k=1,\dots,D$.  For a given region $R$, these tensors are arranged on
a 2D grid as shown in Fig.~\ref{fig:peps}(a).  Adjacent virtual indices
$\mu_k$ in the bulk are contracted (i.e., identified and summed over),
while the ``open'' virtual indices at the boundary are set to
$\bm\alpha\equiv(\alpha_1,\dots,\alpha_{|\partial R|})$. 
One remains with a tensor $c_{i_1,\dots,
i_N}(\bm\alpha)$, which describes a physical state (a PEPS)
$\ket{\Phi_{\bm\alpha}}=\sum c_{i_1,\dots,
i_N}(\bm\alpha)\ket{i_1,\dots,i_N}$.  This defines a linear map
$\mathcal X:|\bm\alpha)\mapsto\mathcal X|\bm\alpha)\equiv
\ket{\Phi_{\bm\alpha}}$ between states $|\bm\alpha)\in (\mathbb
C^D)^{\otimes |\partial R|}$ on the boundary and the subspace $\mathcal
S\equiv \mathrm{span}\,\big\{\ket{\Phi_{\bm\alpha}}\big\}\subset
(\mathbb C^d)^{\otimes |R|}$ of physical states. 
 [We use $|\,\cdot\,)$ to denote
states on the virtual boundary.] Note that equivalently, one can construct
$\ket{\Phi_{\bm\alpha}}$ by placing virtual \emph{bonds}
$\sum_{\mu=1}^D|\mu,\mu)$ with \emph{bond dimension} $D$ along
the edges, the state $|\bm\alpha)$ at the boundary, and 
applying the linear map described by $A$ at every
site~\cite{verstraete:mbc-peps}. 

Having defined the PEPS states $\ket{\Phi_{\bm\alpha}}$ and the PEPS
subspace $\mathcal S=\mathrm{span}\,\{\ket{\Phi_{\bm\alpha}}\}$, let us
now turn towards Hamiltonians for PEPS models. A \emph{parent Hamiltonian}
is a local Hamiltonian $H=\sum h$ such that for any (sufficiently large)
region $R$ \emph{(i)} $h\ge0$, and $H\ket{\Phi_{\bm\alpha}}=0\
\forall\,|\bm\alpha)$, i.e., $H$
is \emph{frustration free} and all states in $\mathcal S$ are ground
states of $H$; and \emph{(ii)} \emph{all} ground states of $H$ are of the
form $\ket{\Phi_{\bm\alpha}}$, $\mathrm{ker}\,H=\mathcal S$; this is
known as the \emph{intersection
property}~\cite{aklt,perez-garcia:parent-ham-2d,schuch:peps-sym}. Given a
PEPS, a parent Hamiltonian can be constructed by choosing $\ker h=\mathcal
S$ for some small region (e.g.\ as a projector), where appropriate
conditions on $A$ (which hold for generic tensors) ensure the intersection
property~\cite{perez-garcia:parent-ham-2d,schuch:peps-sym}; since
$\mathrm{rank}\,\mathcal S\le D^{|\partial R|}$ for large enough $R$, such
$h$ always exist.
The paradigmatic example of a PEPS model is the AKLT model~\cite{aklt},
which is constructed by placing spin-$\tfrac12$ singlet bonds along the edges
and subsequently projecting onto the maximal spin subspace ($S=2$ on the
square lattice); the parent Hamiltonian is obtained by
observing that for any two adjacent sites, the total spin cannot be $S=4$,
and choosing $h=\Pi_{S=4}$ (the projector onto the $S=4$ subspace).

We now start from a PEPS model, specified by $H=\sum h$ and a tensor $A$
characterizing its ground space $\mathcal S$, with a gap $\Delta$ above
the ground space, and consider an arbitrary perturbation to this model,
$H'=H+V=\sum(h+v)$, where $\|V\|\ll\|H\|$. What is the low-energy physics
of the perturbed model $H+V$? In leading order, it is given by the
effective Hamiltonian $H_\mathrm{eff}=\Pi_\mathcal S V \Pi_\mathcal S$,
where $\Pi_\mathcal S$ is the projector onto the ground space $\mathcal S$
of $H$, i.e., the low-energy physics takes place in the subspace $\mathcal
S$.  Since $\mathcal S=\mathrm{span}\,\{\ket{\Phi_{\bm\alpha}}\}$, this
implies that the states which describe the low-energy physics are in
one-to-one correspondence with states $|\bm\alpha)$ on the virtual edge
(via the inverse of the map $\mathcal X$), and thus, the low-energy states
exhibit a 1D structure which is associated to the edge.  Even
more, if the system does not break local symmetries (more technically, if
it satisfies the weak LTQO
condition~\cite{michalakis:local-tqo-ffree,cirac:itb}), these states are
exponentially localized at the edge, i.e., different
$\ket{\Phi_{\bm\alpha}}$ do not differ in the bulk.  Together, this shows
that the low-energy Hamiltonian $H_\mathrm{eff}$ can indeed be understood
as a 1D Hamiltonian acting on degrees of freedom localized at the edge. 

Let us now show how to determine the 1D model which describes the
effective low-energy physics. To this end, we work in the 1D
basis $|\bm\alpha)$ which lives on the virtual edge indices. There, the
perturbation induces a term $(\bm\alpha'|\mathcal
M|\bm\alpha)=\bra{\Phi_{\bm\alpha'}}V\ket{\Phi_{\bm\alpha}}$; this is,
$\mathcal M$ is obtained by sandwiching the Hamiltonian
between a ket PEPS and a bra PEPS, as shown in Fig.~\ref{fig:peps}b.
However, the map $\mathcal X:|\bm\alpha)\mapsto\ket{\Phi_{\bm\alpha}}$ does
not preserve orthogonality, and thus, in order to obtain an edge
Hamiltonian $\mathcal H$ which is isomorphic to $H_\mathrm{eff}$, we need
to orthogonalize $\mathcal M$, $\mathcal H = \mathcal P^{-1} \mathcal M
\mathcal P^{-1}$, where $\mathcal P=\sqrt{\mathcal Q}$,
$(\bm\alpha'|\mathcal
Q|\bm\alpha)=\bra{\Phi_{\bm\alpha'}}\Phi_{\bm\alpha}\rangle$.
Put more formally, we can write $\mathcal X=\mathcal
W\mathcal P$, with $\mathcal P$ a positive map acting on the virtual
indices and $\mathcal W$ an isometry from the virtual to the physical
system; then, the edge Hamiltonian is $\mathcal H = \mathcal
W^\dagger V \mathcal W = \mathcal P^{-1}\mathcal X^\dagger V
\mathcal X \mathcal P^{-1}$ (where $\mathcal X^\dagger V \mathcal X$ is
the tensor network in Fig.~\ref{fig:peps}b), and thus indeed
isomorphic to $H_\mathrm{eff}$.

\begin{figure}[t]
\includegraphics[width=\columnwidth]{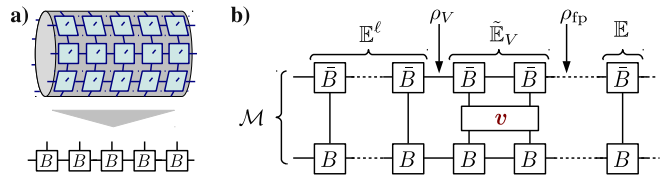}
\caption{\label{fig:bnd-comp}
\textbf{a)} By blocking columns, a PEPS on a cylinder can be mapped to an
MPS. \textbf{b)} Computation of the edge Hamiltonian on a long cylinder,
cf.\ text.  }
\end{figure}

In order to numerically study edge Hamiltonians, we restrict to
translationally invariant PEPS on infinite cylinders.  In this case, we
can block columns and treat the system as an effective 1D
PEPS, i.e., a Matrix Product State (MPS), see Fig.~\ref{fig:bnd-comp}a.
The central object encoding the behavior of the system is the transfer
operator $\mathbb E_O=\sum_{ij} \bra{j}O\ket{i} B^i \otimes \bar B^j$, where
$i$ and $j$ ($B$, $O$) are blocked indices (tensors, operators) for one
column.  We first focus on systems where $\mathbb E\equiv
\mathbb{E}_{\openone}$ has a non-degenerate largest eigenvalue with a gap
below (this corresponds to a unique ground
state with PBC); we assume the largest eigenvalue to
be normalized to $1$.  One first determines the fixed point
$\rho_\mathrm{fp}$ of $\mathbb E$.  Second, one applies
$\rho_V=\tilde{\mathbb E}_V\cdot\rho_\mathrm{fp}$, where $\tilde{\mathbb
E}_V$ is a transfer operator containing one unit cell of $V$ [e.g.\ two
columns for a nearest neighbor (NN) Hamiltonian]. Finally, one iteratively
computes the fixed point $\mathcal M=(\openone+\mathbb E + \mathbb E^2 +
\dots)\rho_V$.  Now, to obtain $\mathcal H$ one needs to orthogonalize
with $\mathcal P=\sqrt{\rho_\mathrm{fp}}$, i.e., the edge Hamiltonian
$\mathcal H$ is obtained as $\mathcal H=\rho_\mathrm{fp}^{-1/2} \mathcal M
\rho_\mathrm{fp}^{-1/2}$; the invertibility of $\rho_\mathrm{fp}$ follows
from the uniqueness of the fixed point of $\mathbb
E$~\cite{perez-garcia:mps-reps}.  The procedure is illustrated in
Fig.~\ref{fig:bnd-comp}b; note that $\mathbb E$ is evaluated as a Matrix
Product Operator and thus, we are limited by the dimension of its
eigenvectors rather than that of $\mathbb E$ itself.  Note also that the
dynamics on the two edges is independent; had we considered a finite
cylinder instead, the dynamics of the two boundaries would be weakly
coupled, with the length scale set by the gap of $\mathbb E$.

An essential point to note about the structure of the edge Hamiltonian is
that it inherits all (on-site) symmetries shared by the PEPS and the
bulk perturbation: Any symmetry action on a PEPS can be moved from the
physical index to an action of the same symmetry on the virtual
indices~\cite{perez-garcia:inj-peps-syms}, and thus ultimately any
symmetry of the state shared by $V$ shows up as a symmetry at the boundary
degrees of freedom and thus in $\mathcal H$; the argument generalizes to
other symmetries such as reflection or time reversal.

\begin{figure}
\includegraphics[width=\columnwidth]{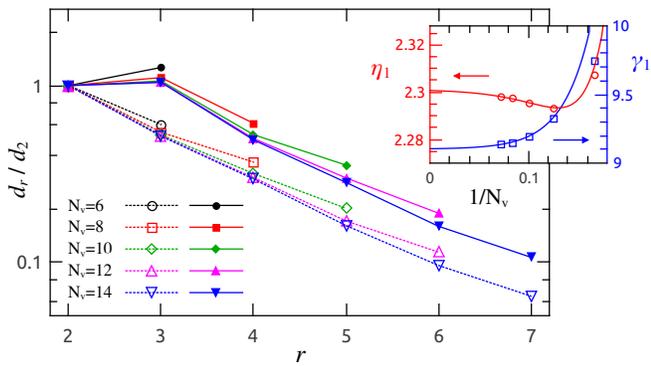}
\caption{\label{fig:aklt-edge}
Edge Hamiltonian for the perturbed AKLT model: Exponential decay of the
interaction strength of range-$r$ terms, $d_r$, with distance, for
different $N_v$, for $\mathcal H_J$ (solid lines) and $\mathcal H_g$
(dotted lines). Inset: Finite size scaling of $\eta_1$ (red circles) and
$\gamma_1$ (blue squares) vs.\ $1/N_v$.
}
\end{figure}

As an example, we have studied the edge Hamiltonian $\mathcal H$
of the square lattice AKLT model on an infinitely long cylinder of
diameter $N_v$, for the class of $\mathrm{U(1)}$ invariant
perturbations
\begin{equation}
    \label{eq:aklt-pert}
V= \sum_{\left\langle ij\right\rangle
}\left[J\, \bm S_{i}\cdot \bm S_{j} 
+g S_{i}^{z}S_{j}^{z}\right] +h\sum_{i}S_{i}^{z}\ ,
\end{equation}
i.e., an anisotropic Heisenberg Hamiltonian with a magnetic field.  Since
$\mathcal H$ is linear in $V$, we can write $\mathcal H = J \mathcal H_J +
g\mathcal H_g + h \mathcal H_h$; as $D=2$, the $\mathcal H_\bullet$
($\bullet=J,g,h$) are spin-$\tfrac12$ Hamiltonians with translational and
$\mathrm{U(1)}$ symmetry [$\mathrm{SU}(2)$ for $\mathcal H_J$]. Note that
due to symmetry, $\mathcal H_J$ is completely determined by $\mathcal
H_g$.

First, let us see whether the $\mathcal H_\bullet$ are sums of local
terms.  To this end, we decompose $\mathcal H_\bullet$ in a Pauli basis,
and denote by $d_r$ the total weight of all terms which span $r$ 
contiguous sites
(see~\cite{cirac:peps-boundaries,schuch:topo-top}).
Fig.~\ref{fig:aklt-edge} shows the result for $\mathcal H_J$ and
$\mathcal H_g$: In both cases $d_r$ decays exponentially with
$r$, indicating that the edge Hamiltonian is approximately local.  Let us
now have a closer look at the individual terms. For $\mathcal H_J$,
symmetries restrict the possible two- and three-body terms 
to Heisenberg couplings, which---following Fig.~\ref{fig:aklt-edge}---are
the dominating terms. More generally, we find
\begin{equation}
\label{eq:HJ-expansion}
\mathcal H_J \approx 
    \sum_{\ell\ge1} \eta_\ell \sum_{i} \bm S_i\cdot\bm S_{i+\ell} 
\end{equation}
where $\eta_1\approx2.298$ and $\eta_2 \approx -2.394$, larger
$\eta_\ell$ decay exponentially, and many-body terms are strongly
supressed.  Remarkably, the NN and next-nearest neighbor (NNN) Heisenberg
terms in $\mathcal H_J$ have essentially the same strength, but opposite
sign (this staggering repeats in the---exponentially
decaying---longer-range $\eta_\ell$ and arises from the alternating parity
of singlets connecting the bulk perturbation to the boundary).
Adding an $S^z_i S^z_{j}$ anisotropy in the bulk leads to an anisotropy
at the edge with a similarly staggered structure and a
renormalized Heisenberg term,
\begin{equation}
\label{eq:Hg-expansion}
\mathcal H_g \approx  \sum_{\ell\ge1}
\left[
    \gamma_\ell \sum_i S_i^zS_{i+\ell}^z + 
    \frac{\eta_\ell-\gamma_\ell}{3} \sum_i \bm S_i\cdot\bm S_{i+\ell}
\right]
\end{equation}
but with supressed NNN amplitudes 
$\gamma_1\approx9.137$, $\gamma_2\approx -4.493$. (The dependence between
the coefficients is due to symmetries.)
Finally, a local magnetic field induces exactly a field of identical
strength at the boundary, $\mathcal H_h=\sum S^z_{i}$, as can be
shown analytically based on symmetries of the state~\footnote{
Applying $S_z$ on the physical index of $A$ translates to applying the sum
of $S_z$ to the virtual indices.  When applying $\sum S_z^i$, the sum of the
$S_z$ appearing at the two ends of each singlet cancel, and one is
left with the $S_z$ at the boundary. Thus, $\mathcal M= (\sum
S_z^i)\rho_\mathrm{fp}= \rho_\mathrm{fp}(\sum S_z^i)$, and the
claim follows. The same argument holds for the response of arbitrary
models with on-site symmetries to local fields.  
}.

\begin{figure}[t]
\includegraphics[width=0.9\columnwidth]{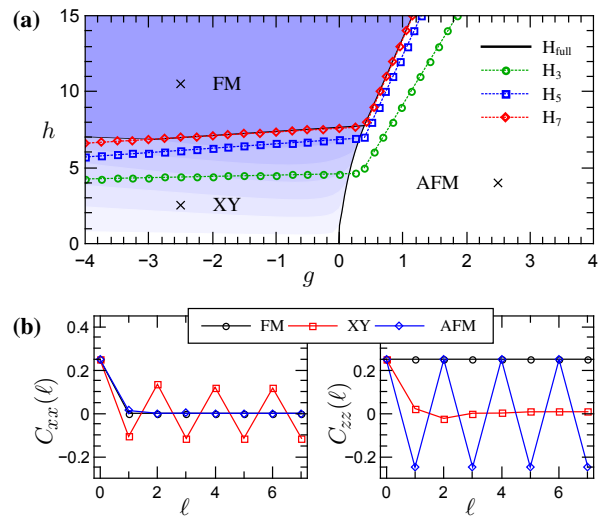}
\caption{\label{fig:phase-diag}
Edge Hamiltonian for the perturbed AKLT model, Eq.~(\ref{eq:aklt-pert}).
\textbf{(a)} Phase diagram as a function of anisotropy $g$ and field $h$,
for $J=1$.  Three phases are observed: a fully polarized ferromagnetic
(FM) phase (with magnetization $m_z=\tfrac12$), an antiferromagnetic (AFM)
phase ($m_{z}=0$), and an XY Luttinger liquid phase.  The shading shows 
$m_z$ for the ground state of the full edge Hamiltonian $\mathcal H$ for
$N_v=14$; the solid lines give phase boundaries determined analytically
using fully polarized and mean-field AFM ansatzes, both for $\mathcal H$
and Hamiltonians $\mathcal H_k$ where the sum in
(\ref{eq:HJ-expansion}) and (\ref{eq:Hg-expansion}) is restricted to
$\ell< k$.  \textbf{(b)} Correlation functions
$C_{xx}(\ell)=\left\langle S_{i}^{x}S_{i+\ell}^{x} \right\rangle$ and
$C_{zz}(\ell)=\left\langle S_{i}^{z}S_{i+\ell}^{z} \right\rangle$ for the three
phases, computed at the points marked $\times$ in (a). DMRG calculations
for $\mathcal H_k$ show that in the XY phase, $C_{xx}$ decays
algebraically.} \end{figure}

We have studied the phase diagram of the edge for $J>0$ using exact
diagonalization supplemented by DMRG and analytical arguments, see
Fig.~\ref{fig:phase-diag}.  We find that the model exhibits three
phases---a fully polarized ferromagnetic phase, an antiferromagnetic
phase, and an XY Luttinger liquid phase.  By choosing the appropriate bulk
perturbation $V$, we can thus achieve either gapped, gapless, or symmetry
broken phases at the edge, and induce phase transitions between them.

A natural question to ask at this point is whether we can achieve any edge
Hamiltonian $\mathcal H$ we want.  Since for a trivial bulk phase the
mapping $\mathcal X$ from the edge to the bulk is injective, the answer
there is indeed yes.  Even more, any local $\mathcal H$ can be obtained from an
approximately local bulk perturbation: At the RG fixed point where $A$ is
the identity, this is clear.  Now for any model connected to it via a gapped
path, we can obtain its ground space via a quasi-adiabatic
evolution~\cite{hastings:quasi-adiabatic} of the original ground space;
the correspondingly evolved bulk perturbation is then
quasi-local and yields the desired $\mathcal H$.  Thus, we find that for a
trivial bulk phase, the edge is never protected~\footnote{Note that it is
still possible to protect edge properties by symmetries in the
bulk~\cite{gu:spt-fermions,chen:spt-bosons}---e.g., in the AKLT model,
$\mathrm{SU}(2)$ symmetry rules out a gapped edge which does not break any
symmetry~\cite{lieb:LSM}.}.

We now turn towards topologically ordered systems, and investigate whether the
bulk order can protect the physics at the edge.  In these systems, the
PEPS tensor is invariant under a symmetry action on the virtual indices
which can be identified with particle types (charges) $p$ of the
topological model.  Therefore, the transfer operator $\mathbb E$ of a
column,
Fig.~\ref{fig:bnd-comp}, is degenerate, with its maximal eigenvectors
$\rho_{\mathrm{fp}}^p$ being supported on the sector with total
topological charge $p$~\cite{schuch:topo-top}.  In particular, the fixed
point $\mathbb E^\infty$ in Fig.~\ref{fig:bnd-comp}b is of the form
$\mathbb E^\infty = \sum_p \ket{\rho_{\mathrm{fp},L}^p}
\bra{\rho_{\mathrm{fp},R}^{p^*}}$, with $p^*$ the anti-particle of $p$.

Let us now for a moment fix $p$ in the sum: Then, we are essentially back in
the scenario which we had for non-degenerate $\mathbb E$, in that any
perturbation induces an effective edge Hamiltonian on the two edges
independently.  However, there is an important difference:
$\rho_{\mathrm{fp},L}^p$ does not have full rank, but is supported on the
sector with topological charge $p$.  Thus, only boundary conditions
$|\bm\alpha)$ in this sector will correspond to a non-zero physical state
$\ket{\Phi_{\bm\alpha}}$ and thus to admissible gapless excitations. At
the same time, the label $p$ is also preserved by $\mathbb E_V$ (since it
emerges from a symmetry acting solely on the virtual indices of the PEPS
tensor~\cite{schuch:peps-sym}), and thus, $\mathcal M_L^p$ is
also supported only in that sector. Thus, we can still orthogonalize it
using the pseudoinverse of $(\rho_{\mathrm{fp},L}^p)^{1/2}$, and obtain an
effective edge Hamiltonian $\mathcal H_L^p$ for the sector with charge
$p$; analogously, we obtain an edge Hamiltonian $\mathcal H_R^{p^*}$ for
the right edge.

The full edge Hamiltonian is now obtained by putting both edges
together and summing over $p$; it is of the form
\[
\mathcal H = \Pi_0 \, (\mathcal H_L\otimes \openone_R + 
    \openone_L\otimes \mathcal H_R)\,\Pi_0\ ,
\]
where $\mathcal H_{L,R}=\sum_p \mathcal H_{L,R}^p$, and $\Pi_0$ is the
projector onto the sector with total charge $p=0$ for both boundaries
together. This implies that the edge Hamiltonian for a single edge must
conserve the topological charge; this edge symmetry is protected by the
topological order in the bulk and can stabilize non-trivial properties of
the edge Hamiltonian~\cite{gu:spt-fermions}.  Let us illustrate this for
the Toric Code (TC)~\cite{kitaev:toriccode}, where the spin-$\tfrac12$
edge Hamiltonian is constrained by a quasi-fermionic $\mathbb Z_2$ parity
superselection rule.  Since the TC is an RG fixed point, there is a
one-to-one local unitary correspondence between virtual and physical
degrees of freedom at the edge up to the parity
constraint~\cite{schuch:peps-sym}, allowing to engineer any
parity-preserving edge Hamiltonian.  In
particular, $V=-\sum_{\langle ij\rangle}S^x_iS^x_j$ yields Ising models
$\mathcal H_L=\mathcal H_R=-\sum S^x_iS^x_{i+1}$ at the edges, whose even
and odd parity ground states are the GHZ states
$\ket{+\ldots+}\pm\ket{-\ldots-}$. Thus, each of the edges is an Ising
model in a GHZ state---a macroscopic superposition---which is
protected against spontaneous symmetry breaking by \emph{arbitrary}
local perturbations, something which is impossible in a conventional 1D
spin system; this is in close analogy to the protection of a fermionic
Majorana chain~\cite{kitaev:majorana-chain}.

We have computed the edge Hamiltonian for the topological 
RVB state on the kagome lattice, which is a $D=3$
PEPS~\cite{verstraete:comp-power-of-peps,schuch:rvb-kagome}, for a bulk
perturbation $V=\sum_{\langle i,j\rangle} \bm S_i\cdot\bm S_j$.  We find that
$\mathcal H_L$ and $\mathcal H_R$ are again approximately local, while the
per-sector Hamiltonians $\mathcal H_{L/R}^p$ are not; the latter 
is due to the fact that $\mathcal H_{L/R}^p$ contain a projector onto a
superselection sector, in direct analogy to what has been found for
the Hamiltonians reproducing the entanglement spectrum in the
case of topological models~\cite{schuch:topo-top}.  The symmetry of the
RVB PEPS strongly restricts the possible local terms, implying that the
structure of the edge is that of a spinful particle or a hole, similar to
a  $t$-$J$ model~\cite{poilblanc:rvb-boundaries};  in that language and
the notation of~\cite{poilblanc:rvb-boundaries},  the leading terms of the
edge Hamiltonian for $N_v=8$ are a NN Heisenberg term ($J_1\approx 0.233$), 
chemical potential ($c_2\approx 0.177$), NN hopping ($t_1\approx -0.158$),
and NN singlet creation ($\Delta_1\approx -0.086$).  We have also
considered a chiral perturbation $V=\sum \bm S_i \cdot (\bm S_j\times \bm
S_k)$, where the sum runs over all triangles, and found that the dominant
term at the edge is given by a chiral current of particles,
$\mathcal H_L \approx \sum ia_{s,k}^\dagger
a^{\phantom\dagger}_{s,k+1}+\mathrm{h.c.}$, carrying $64.5\%$ of the total
weight in $\mathcal H_L$.  Note, however, that such a term by itself does 
not give rise to a protected chiral edge mode. [A similar chiral
perturbation to the AKLT model gives in leading order rise to a chiral
spin current $\mathcal H \approx \sum \bm S_k\cdot(\bm S_{k+1}\times\bm
S_{k+2})$; this is the simplest $\mathrm{SU}(2)$ invariant
spin-$\tfrac12$ Hamiltonian.]

In this paper, we have studied edge theories in the framework of
PEPS models.  We have demonstrated that the effective low-energy theory
lives on the virtual degrees of freedom at the boundary, which allows to
explicitly obtain the edge Hamiltonian in the vicinity of these models.  In
the trivial phase, this allows to engineer arbitrary edge Hamiltonians,
while topological bulk phases carry symmetries at their boundary which can
protect the physics at the edge. Thus, protected physics at the edge is a
signature of topological order in the bulk, and we expect that one can
characterize the type of bulk topological order from the the protected
properties of the edge~\cite{inprep}.  All results equally apply to
fermionic systems~\cite{kraus:fPEPS}.  While we focused on a perturbative
regime around PEPS models, we expect our findings to apply more generally:
First, PEPS approximate ground states of local Hamiltonians
well~\cite{hastings:mps-entropy-ent} and any (generic) PEPS has a parent
Hamiltonian associated with
it~\cite{perez-garcia:parent-ham-2d,schuch:peps-sym}, suggesting that many
systems have a PEPS model closeby; and second, the identification of the
low-energy physics with the virtual degrees of freedom at the edge extends
to any system connected to a PEPS model by a gapped path, by
quasi-adiabatical evolution of the ground
space~\cite{hastings:quasi-adiabatic}.

One question left open is the possible correspondence between entanglement
spectrum and edge
physics~\cite{li:es-qhe-sphere,fidkowski:freeferm-bulk-boundary,qi:bulk-boundary-duality}
beyond that emerging from their joint symmetry structure.  E.g., for the
RVB model the Heisenberg term in $\mathcal H$ is much enhanced as compared
to the Hamiltonian derived from the entanglement
spectrum~\cite{cirac:peps-boundaries,poilblanc:rvb-boundaries}, this can 
be seen as a trace of the Heisenberg bulk perturbation. It would be
interesting to study this further by applying our framework to frustrated
PEPS models (such as variationally minimized iPEPS) which exhibit edge
dynamics without perturbations.

We acknowledge helpful discussions with Z.-C.~Gu, D.~P\'e{}rez-Garc\'\i{}a
and J.~Preskill.  Parts of this work were done at the Simons Institute
for the Theory of Computing in Berkeley, and at the
Centro de Ciencias de Benasque
Pedro Pascual in Benasque, Spain.   N.S.\ and L.L.\ acknowledge support by
the Alexander von Humboldt foundation and the EU project QALGO.  D.P.\
acknowledges partial support by the Agence Nationale de la Recherche under
grant No.~ANR~2010~BLANC~0406-0.  J.I.C.\ and F.V.\ acknowledge support by
the EU project SIQS.  F.V.\ acknowledges support from the FWF (Foqus and
Vicom), the ERC (QUERG), and the FWO (Odysseus grant).

\end{document}